\documentclass[11pt]{article}
\usepackage{amsfonts}
\usepackage{amsbsy} 
\usepackage{epsfig}
\usepackage{latexsym}
\usepackage{graphicx, tikz}
\usetikzlibrary{shapes.geometric, arrows}
\usepackage{amsmath, mathrsfs}
\usepackage{subcaption}
\usepackage{breqn}
\usepackage[style=numeric-comp, backref=true, url=false, sorting=none, maxbibnames=9]{biblatex}
\usepackage[hidelinks]{hyperref}
\usepackage{setspace} 
\usepackage[nottoc,numbib]{tocbibind}
\usepackage[margin=0.8in]{geometry}
\usepackage{cancel, soul}

\newcommand{\e}{\mathrm{e}}
\newcommand{\im}{\mathrm{i}}
\renewcommand{\d}{\mathrm{d}}

\begin{document}
\hfuzz=10pt
\vfuzz=10pt
\title{{\Large \bf{Near-Horizon Symmetries of Local Black Holes in General Relativity}}}
\author{M M Akbar\footnote{akbar@utdallas.edu} \footnote{Corresponding Author} $^a$
\ \,\,\&
S M Modumudi\footnote{saimadhav.modumudi@utdallas.edu} $^b$
\\
\\
$^a$ Department of Mathematical Sciences,
\\ University of Texas at Dallas, 
\\ 800 W Campbell Rd,
\\ Richardson, Texas, USA.
\\
\\
$^b$ Department of Physics,
\\ University of Texas at Dallas, 
\\ 800 W Campbell Rd,
\\ Richardson, Texas, USA.}
\maketitle

\begin{abstract}
We analyze the near-horizon symmetries of static, axisymmetric, four-dimensional black holes with spherical and toroidal horizon topologies in vacuum general relativity. These black hole solutions, collectively referred to as local/distorted black holes, are known in closed form and are not asymptotically flat. Building on earlier works in the literature that primarily focused on black holes with spherical topology, we compute the algebra of the Killing vector fields that preserve the asymptotic structure near the horizons and the algebra of the associated Noether-Wald charges under the boundary conditions that produce the spin-$s$ BMS$_d$ and the Heisenberg-like algebras. We show that a similar analysis extends to all local axisymmetric black holes. The toroidal topology of the holes changes the algebras considerably. For example, one obtains two copies of spin-$s$ BMS$_3$ instead of spin-$s$ BMS$_4$. We also revisit the thermodynamics of black holes under these boundary conditions. While previous studies suggested that spin-$s$ BMS$_d$ preserves the first law of thermodynamics for isolated horizons ($\kappa= \text{const.}$), our analysis indicates that this is not generally the case when the spin parameter $s$ is nonzero. A nonzero $s$ can be seen as introducing a conical singularity (in the Euclidean quantum gravity sense) or a Hamiltonian that causes soft hairs to contribute to the energy. This leads us to interpret the spin-$s$ BMS$_d$ boundary condition as arising in the context of dynamical black holes.
\end{abstract}

\section{Introduction}
In 1962, Bondi, van der Burg, Metzner, and Sachs, while analyzing gravitational waves in asymptotically flat spacetimes, discovered that the coordinate transformations that preserve the asymptotic structure of the metric near future null infinity $(\mathscr{I}^+)$ formed a symmetry group much larger than the Poincar\'e group, now known as the Bondi-Metzner-Sachs (BMS) group \cite{bondi1962, sachs1962_waves, sachs1962_asym}. It is a semi-direct product of the Lorentz group and an infinite-dimensional group of transformations called ``supertranslations". Around the same time, Penrose, Newman, Geroch, and others developed a topological approach to asymptotically flat spacetimes \cite{Penrose:1962, Penrose:1965, Newman-Penrose-1968, Geroch1977}. These efforts allowed the BMS group to be reinterpreted as the exact global diffeomorphism group of $\mathscr{I}^+$ (after quotienting out the trivial diffeomorphisms, see, for example, \cite{ashtekar2014}). On the quantum side, since the BMS group only admitted discrete-spin (irreducible) representations, in contrast to the Poincar\'e group which also allows continuous-spin representations, it offered an explanation for why nature favors discrete-spin particles, including the spin-$\frac{1}{2}$ particles \cite{Newman1965,  mccarthy1972_structure, McCarthy-74}.

After decades of relative obscurity, the BMS group has experienced a resurgence of interest in recent years, driven by the equivalences between gravitational memory effect, soft theorems, and asymptotic symmetries \cite{Strominger-Zhiboedov:2014}. It has also been found that the Lorentz part of the BMS group can be expanded to a larger group of ``superrotations" \cite{Barnich_2011}. The associated  Noether charges form a Witt algebra, a special case of the Virasoro algebra with no central extension, while the supertranslations produce an Abelian current \cite{Barnich_2011}. Generally, by ``BMS group" these days, we mean the group including superrotations.

In 2015, Hawking realized that supertranslations near the horizons of stationary black holes could provide a possible resolution to the information loss paradox \cite{Hawking:2015} and the zero-energy conserved charges or ``soft hairs" associated with these symmetries could encode information on the black hole horizons, as a form of holography \cite{Hawking:2016a, Hawking:2016}. Subsequently, a number of authors have studied the near-horizon symmetries of various black holes and worked out the associated algebras \cite{Donnay:2015, Donnay:2016, Donnay:2019, Grumiller-2020, Anabalon2021, Giribet2023}. In particular, with suitable fall-off conditions, the symmetry algebra (for spherical black holes) was found to be a semi-direct sum of two copies of the Witt algebra and an Abelian current algebra with structure constants slightly different from those of the BMS algebra \cite{Donnay:2015, Donnay:2016}. Considering the fall-off conditions in a $3+1$ Arnowitt-Deser-Misner (ADM) decomposition, it was subsequently shown that one can obtain ``spin-$s$ BMS$_d$ algebra" for a certain choice of boundary condition parametrized by a parameter $s$ \cite{Grumiller-2020}. This algebra gives the familiar BMS$_d$ algebra for $s=1$\footnote{In Figure \ref{fig:DifferentBMS}, we show the different BMS groups and the relations among themselves and the Poincar\'e group.}. For another boundary condition, one obtains a Heisenberg-like algebra, the generators of which reproduce the generators of the BMS-like algebra under composition \cite{Afshar:2017, Grumiller-2020}. { Similar infinite-dimensional algebras are also found near generic null hypersurfaces (see, for example, \cite{Adami_2020a, Adami_2020b, Adami_2022})}. 

In this paper, we follow the above approaches to study the near-horizon symmetries of static, axisymmetric (Weyl) black holes in vacuum general relativity. These black holes, distorted by matter sources at a distance, are not asymptotically flat (except for the special case of the Schwarzschild solution) and can have spherical or toroidal horizon topologies despite being vacuum solutions \cite{stephani_exact_sol}. In their seminal work, Geroch and Hartle obtained these solutions in closed form and referred to them as ``local black holes'' \cite{GerochHartle}. Chandrasekhar later re-derived the spherical topology solutions in Schwarzschild-like coordinates, demonstrating how these can be seen as distortions of the Schwarzschild metric by distant multipole sources \cite{Doroshkevich-Zeldovich-65, chandrasekharbook}. Xanthopoulos extended Chandrasekhar's treatment to toroidal holes \cite{LocalToroidalXanthapolous}. 

In Section \ref{sec:gen-treatment}, we review the near-horizon formalism for spherical black holes in Gaussian coordinates.
In Section \ref{sec:local-black-holes}, we introduce axisymmetric vacuum black holes, and derive their near-horizon symmetries. We obtain exact expressions for some of the charges and the algebra of their generators and note how the algebras change fundamentally for the toroidal case. In Section \ref{sec:thermodynamics}, we study thermodynamics for the spin-$s$ BMS$_d$ boundary condition and show that for a nonzero spin parameter one does not get the first law of thermodynamics for isolated black holes. The first law holds for $s=0$ (when the algebra is two copies of Witt mentioned above), and for the Heisenberg-like boundary condition.

\section{General Treatment}\label{sec:gen-treatment}
In studies of asymptotic symmetries in general relativity, null coordinates are often employed. While earlier works on near-horizon symmetries utilized null coordinates \cite{Donnay:2015, Donnay:2016}, finding them in closed form is challenging for non-asymptotically flat spacetimes or horizons with non-spherical topology, such as those we consider in this paper. We thus adopt Gaussian-type coordinates (that is, $\delta_{\mu\rho}$ to first order) as used in later works \cite{Donnay:2019, Giribet2023, Grumiller-2020} --- these works take inspiration from the ``membrane paradigm'' of black hole horizons pioneered by Price and Thorne \cite{PriceThorne}. We would like to add here that some earlier works also considered near-horizon symmetries in the context of horizon microstates \cite{Carlip_1995, Carlip_1999, Hotta_2001, Carlip:2002}.

In this section, we summarize the results and approaches adopted in \cite{Donnay:2016, Grumiller-2020}, presenting the necessary details in a self-contained manner for use in the subsequent sections. In either approach, one begins by finding a suitable radial coordinate and writes the black-hole geometry as a product of horizon geometry and a codimension-two Minkowski metric in Rindler coordinates. This is similar to the approach one takes in Euclidean quantum gravity where, in addition, one periodically identifies the Euclideanized time, and the periodicity needed to obtain a non-conical two-dimensional Euclidean space gives the temperature of the hole (see, for example, \cite{Gibbons-Hawking-Book}). If $\rho$ is the suitable radial coordinate such that the horizon is located at $\rho=0$,
one can impose the following boundary conditions \cite{Grumiller-2020}\footnote{These fall-off conditions are a bit more general than those used in \cite{Donnay:2016}. However, the Lie algebra of the Killing vectors will not differ between the two fall-offs.}
\begin{equation}\label{eq:Fall-off}
\begin{aligned}
g_{tt} &= -\kappa^2 \rho^2 + \mathcal{O}(\rho^3)~, & g_{\rho\rho} &= 1 + \mathcal{O}(\rho)~, & g_{t\rho} &= \mathcal{O}(\rho^2) \\
g_{tA} &= R_A (\phi^B) \rho^2 + \mathcal{O}(\rho^3)~, & g_{\rho A} &= M_{A}(\phi^B)\rho + \mathcal{O}(\rho^2)~, & g_{AB} &= \Omega_{AB}(\phi^D) + \mathcal{O}(\rho^2) ~,
\end{aligned}
\end{equation}
where $R_A$, $M_A$ and $\Omega_{AB}$ are functions of their arguments, $\kappa = -\frac{1}{2} \nabla^\mu \zeta^\nu \nabla_\mu \zeta_\nu$ is the surface gravity with $\zeta$ being the Killing vector on the horizon, $\phi^A$ are angular coordinates with $A \in \{2,3\}$.

The Killing vectors preserving the above asymptotic expansions work out to be
\begin{equation} \label{eq:Killing-Vectors}
\xi^t = \frac{1}{\kappa} T(t, \phi^A) + \mathcal{O}(\rho)~, \qquad \xi^\rho = \mathcal{O}(\rho^2)~, \qquad \xi^A = \Phi^A(\phi^B) + \mathcal{O}(\rho^2),
\end{equation}
where $T$ and $\Phi^A$ are related via $\partial_t T + \Phi^A \partial_A \kappa = \delta\kappa$ (which follows from the fall-off conditions on Lie derivatives of the metric, see equation (2.12) in \cite{Donnay:2016}), but are otherwise arbitrary functions of their arguments. For constant $\kappa$, this equation implies that $T$ is only a function of the angular coordinates. In any case, $T$ and  $\Phi^A$ represent supertranslations and superrotations, respectively.

\subsection*{Algebra}
In general, there are two ways to approach the symmetry algebra: the algebra of the Killing vectors using (modified) Lie brackets and the algebra of the associated conserved charges using Poisson brackets. 

\bigskip\noindent \textbf{Killing-Vector Algebra:} To find the algebra of the Killing vectors, one makes use of the modified Lie bracket \cite{Barnich-Troessaert}
\begin{equation}\label{eq:LieBracket}
[\xi_1, \xi_2] = \mathcal{L}_{\xi_1} \xi_2 - \delta_{\xi_1} \xi_2 + \delta_{\xi_2} \xi_1~,
\end{equation}
where, $\delta_{\xi_i} \xi_j$ represents the change induced in $\xi_j[g]$ by the variation $\mathcal{L}_{\xi_i} g_{\mu\nu}$, that is, $\delta_{\xi_1}\xi_2^\mu = \xi_2^\mu[\mathcal{L}_{\xi_1}g]$   . The (closed) algebra then works out to be
\begin{equation}\label{eq:LieBracket-detail}
[\xi(T_1,\Phi_1^A), \xi(T_2, \Phi_2^A)]^\mu = \xi_1^\nu \partial_\nu \xi_2^\mu - \xi_2^\nu \partial_\nu \xi_1^\mu - \xi_2^\mu[\mathcal{L}_{\xi_1}g] + \xi_1^\mu[\mathcal{L}_{\xi_2}g]~,
\end{equation}
which can be written as $\xi(T_{12}, \Phi_{12}^A)$ where,
\begin{equation}
\begin{aligned}
\frac{T_{12}}{\kappa} &= \frac{T_1}{\kappa} \partial_t\left( \frac{T_2}{\kappa}\right) - \frac{T_2}{\kappa} \partial_t \left( \frac{T_1}{\kappa} \right) + \Phi^A_1 \partial_A \left(\frac{T_2}{\kappa}\right) - \Phi^A_2 \partial_A \left(\frac{T_1}{\kappa} \right)~, \\
\Phi_{12}^A &= \Phi^B_1 \partial_B \Phi^A_2 - \Phi^B_2 \partial_B \Phi^A_1~.
\end{aligned}
\end{equation}
For spherical black holes and constant surface gravity $\kappa$ the algebra has been worked out \cite{Donnay:2016, Donnay:2019}. One uses the  complex coordinates $\zeta, \Bar{\zeta}$ on the sphere, expands $T(x^{A})$ and $\Phi^A (x^{A})$ in Laurent modes
\begin{equation}\label{eq:function-expansion-spherical}
T = \sum_{mn} \zeta^m \bar{\zeta}^n \tau_{mn} \qquad Y^\zeta = \sum_{m} y_m \zeta^m \qquad Y^{\bar{\zeta}} = \sum_{m} y_m \bar{\zeta}^m~,
\end{equation}
defines a basis for the vector space
\begin{equation}
\mathcal{T}_{mn} = \xi(\zeta^{m+1}\bar{\zeta}^{n+1},0,0) \qquad \mathcal{Y}_{m} = \xi(0,\zeta^{m+1},0) \qquad \bar{\mathcal{Y}}_{n} = \xi(0,0,\bar{\zeta}^{n+1})~,
\end{equation}
and obtains the following nonzero commutation relations 
\begin{equation}
\begin{aligned}\label{WittAbelian}
&[\mathcal{Y}_m, \mathcal{Y}_n] = (n-m) \mathcal{Y}_{m+n}~, \qquad 
& &[\bar{\mathcal{Y}}_m, \bar{\mathcal{Y}}_n] = (n-m) \bar{\mathcal{Y}}_{m+n}~, \\[5pt] 
&[\mathcal{Y}_m, \mathcal{T}_{pq}] = p \mathcal{T}_{m+p~q}~,\qquad
& &[\bar{\mathcal{Y}}_m, \mathcal{T}_{pq}] = n \mathcal{T}_{p~m+q}~.
\end{aligned}
\end{equation}
This is a semi-direct sum of two commuting copies of the Witt algebra generated by $\mathcal{Y}_m$ and $\bar{\mathcal{Y}}_n$ which represent the superrotations, and an Abelian current algebra generated by $\mathcal{T}_{mn}$'s which represent the supertranslations. This, for example, has been verified for various C-metrics at both their black hole and acceleration horizons as well as for the (cosmological) de Sitter horizon \cite{Donnay:2019, Anabalon2021}. One can physically interpret the mode $\mathcal{T}_{00}$ as the generator of rigid translations and therefore can be associated with energy. All other generators, i.e.,~$\mathcal{Y}_m$, $\bar{\mathcal{Y}}_n$ and $\mathcal{T}_{mn}$, commute with $\mathcal{T}_{00}$, and they can thus be generators of soft hair on the horizon \cite{Donnay:2016}.\par

\bigskip\noindent\textbf{Charge Algebras:} For the charge algebra, one starts with the charge variation, which can be obtained from the covariant formalism \cite{Regge_Teitelboim_1974, Wald-Zoupas, Barnich_Brandt_2002}
\begin{equation}\label{eq:gen-charge-variation}
\cancel{\delta}Q_\xi = \frac{1}{16\pi} \int \left(\d^{d-2} x\right)_{\mu\nu} \sqrt{-g} \left[ \xi^{[\nu}\nabla^{\mu]} h - \xi^{[\nu}\nabla_\sigma h^{\mu]\sigma}  + \xi_\sigma \nabla^{[\nu} h^{\mu]\sigma} + \frac{1}{2} h \nabla^{[\nu} \xi^{\mu]} - h^{\sigma[\nu}\nabla_\sigma\xi^{\mu]}\right] 
\end{equation}
where $\left(\d^{d-2} x\right)_{\mu\nu} \equiv \frac{1}{(d-2)!} \varepsilon_{\mu\nu\alpha_1 \alpha_2 \cdots \alpha_{d-2}} \d x^{\alpha_1} \wedge \d x^{\alpha_2} \wedge \cdots \wedge \d x^{\alpha_{d-2}}$, $h_{\mu\nu} \equiv \delta g_{\mu\nu}$, $\xi$ is the asymptotic Killing vector, and the slash on $\delta$ represents that the variation might not be integrable. Using the boundary conditions \eqref{eq:Fall-off}, and the Killing vectors \eqref{eq:Killing-Vectors}, one obtains the following dynamical fields \cite{Grumiller-2020}:
\begin{equation}\label{eq:p-and-j}
\mathcal{P} = \frac{\sqrt{\Omega}}{8\pi} \qquad \text{and} \qquad \mathcal{J}_A = \frac{\sqrt{\Omega}}{16\pi\kappa}\left( \partial_t M_A - R_A \right)~,
\end{equation}
with
\begin{equation}\label{eq:p-j-variation}
\begin{aligned}
\delta\mathcal{P} &= \Phi^A \partial_A \mathcal{P} + \mathcal{P} \partial_A\Phi^A~,\\
\delta\mathcal{J}_A &= \mathcal{P} \partial_A T +  \Phi^B\partial_B \mathcal{J}_A + \mathcal{J}_B\partial_A \Phi^B + \mathcal{J}_A\partial_B\Phi^B~,
\end{aligned}
\end{equation}
where, $M_A$ and $R_A$ are the leading order terms in the expansions for $g_{\rho A}$ and $g_{tA}$, respectively (see \eqref{eq:Fall-off}).
In terms of these dynamical fields, one can write the charge variation \eqref{eq:gen-charge-variation} as
\begin{equation}\label{eq:charge-variation}
\delta Q = \int \d^{d-2}x \left[ T \delta\mathcal{P} + \Phi^A \delta\mathcal{J}_A \right]~.
\end{equation}
The integrability of $\delta Q$ will depend on the boundary conditions imposed. This is integrable, for example, for constant surface gravity \cite{Donnay:2016, Donnay:2019}. The Poisson brackets of the charges in this case work out to be a semi-direct sum of two copies of the Witt and an Abelian current algebra, just the same as their Killing vector algebra above in \eqref{WittAbelian} \cite{Donnay:2015, Donnay:2016, Donnay:2019}. This is very close to the BMS (with superrotations) algebra with slightly different structure constants. Note that the expression for variation of charges in \cite{Donnay:2016, Donnay:2019} (see, for example, equation (2.32) in \cite{Donnay:2016}) looks slightly different from equation \eqref{eq:charge-variation}. However, they are identical expressions in different notations.

\begin{figure}[htb]
    \centering
    \includegraphics[width=0.8\textwidth]{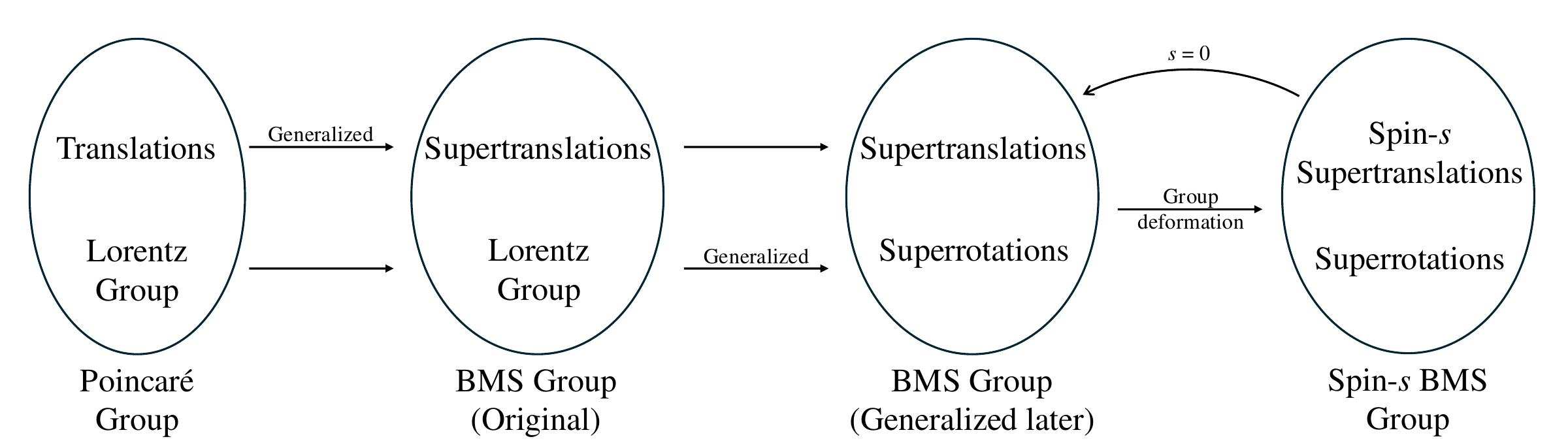}
    \caption{Evolution of the BMS group: Originally discovered by Bondi, Metzner, van der Burg, and Sachs, the BMS group was later generalized in 2011 by Barnich and Troessaert to include superrotations. Various deformations of the BMS algebra were subsequently studied (see \cite{Safari-Jabbari:2019} and references therein).}
    \label{fig:DifferentBMS}
\end{figure}

Considering the problem in a rotating frame within the Arnowitt–Deser–Misner (ADM) framework of general relativity, one can introduce a ``spin parameter" $s$ and impose suitable boundary conditions to ensure that the boundary term --- typically required for a well-defined Hamiltonian formulation --- is integrable. Under these conditions, a deformation of the BMS group, known as the ``spin-$s$ BMS group," emerges \cite{Grumiller-2020} (see Figure \ref{fig:DifferentBMS}). For $s=0$, this recovers the algebra obtained by \cite{Donnay:2016, Donnay:2019}, described above in \eqref{WittAbelian}, while for $s=1$, it yields the BMS$_d$ algebra.\footnote{{ See \cite{Adami_2021b} for surface charges on a generic null boundary.}}\nocite{Safari-Jabbari:2019}

One can then simply drop the deltas and write the integrated charge as
\begin{equation}\label{eq:integrated-charge}
Q = \int \d^{d-2}x \left[ T \mathcal{P} + \Phi^A \mathcal{J}_A \right]~.
\end{equation}
A second set of boundary conditions (without the spin parameter) leads to the Heisenberg algebra for the charge generators \cite{Grumiller-2020}. The boundary conditions for obtaining the BMS and the Heisenberg algebra are detailed below. For more details, see \cite{Grumiller-2020, Grumiller-Jabbari-Book}.
\medskip

\noindent\textit{Boundary Conditions for BMS:} The ADM formalism is characterized by the lapse function $N(x^{\mu})$ and the shift vector $N^i(x^\mu)$. Near a horizon they can be expanded as follows \cite{Grumiller-2020}
\begin{equation}
N= \mathcal{N} \rho + \mathcal{O}(\rho^2),\,\,\,{N}^A=\mathcal{N}^A + \mathcal{O}(\rho^2),\,\,\, {N}^\rho=\mathcal{O}(\rho^2).
\end{equation}
where $\mathcal{N}$ and $\mathcal{N}^A$ are functions of all other coordinates except $\rho$. For these to match with the corotating-frame behavior (\ref{eq:Fall-off}) one must have $\mathcal{N}=\kappa$ and $\mathcal{N}^A=0$. The boundary term takes the form (in the limit of small $\rho$)
\begin{equation}
\delta I_B = -\int \d t \d^{d-2}x \left( \mathcal{N}\delta\mathcal{P} + \mathcal{N}^A \delta\mathcal{J}_A \right)~,
\end{equation}
and its integrability requires the existence of a functional $F$ such that
\begin{equation}
\mathcal{N} = \frac{\delta F}{\delta \mathcal{P}} \qquad \text{and} \qquad \mathcal{N}^A = \frac{\delta F}{\delta \mathcal{J}_A}~.
\end{equation}
When $\mathcal{N}$ and $\mathcal{N}_A$ and the symmetry generators $T$ and $\Phi^A$ have the same field dependence ($\mathcal{N}\to T$, $\mathcal{N}^A \to \Phi^A$), the integrability of the boundary term implies the integrability of the charge variation.

To obtain spin-$s$ BMS algebra, variation of the leading term of the shift vector is kept zero, that is, $\delta\mathcal{N}^A=0$, and the leading order term of the lapse function is allowed to vary as $\mathcal{P}^{s/(d-2)}$ taking $\mathcal{N} = \mathcal{N}^{(s)}\mathcal{P}^{s/(d-2)}$, where $\delta\mathcal{N}^{(s)} = 0$. With this the charge variation can be integrated as
\begin{equation}\label{eq:gen-charge}
Q[T^{(s)}, \Phi^A] = \int \d^{d-2}x \left[ T^{(s)}\mathcal{P}^{(s)} + \Phi^A\mathcal{J}_A \right]~,
\end{equation}
where,  
\begin{equation}\label{eq:t-p-r}
T^{(s)} \equiv T\mathcal{P}^{-r}~, \qquad \mathcal{P}^{(s)} \equiv \frac{\mathcal{P}^{r+1}}{r+1}~,\qquad r\equiv\frac{s}{d-2}.
\end{equation}
One can check that equation (\ref{eq:charge-variation}) is satisfied by $T$ and $\delta Q$ for any $s$. On the other hand, this introduces a denominator to the first term of the integrated charge, equation (\ref{eq:integrated-charge}):
\begin{equation}\label{eq:integrated-chargewithr}
Q = \int \d^{d-2}x \left[ T \frac{\mathcal{P}}{r+1} + \Phi^A \mathcal{J}_A \right]~.  
\end{equation}
This factor is crucial in obtaining the spin-$s$ BMS algebra. We will return to this point in section \ref{sec:thermodynamics} when discussing thermodynamics. The total Hamiltonian is the generator of unit time translations, given by \cite{Grumiller-2020}
\begin{equation}\label{eq:Hamiltonian}
H \equiv Q[\partial_t] = \int \d^{d-2}x \mathcal{N}^{(s)}\mathcal{P}^{(s)}.
\end{equation}
From the definition of $\mathcal{N}^{(s)}$, we can write it as $\mathcal{N}\mathcal{P}^{-r} = \mathcal{N}\left[ (r+1) \mathcal{P}^{(s)} \right]^{-r/(r+1)}$. In a co-rotating frame, $\mathcal{N}$ is $\kappa$, which implies $\mathcal{N}^{(s)} = \kappa \left[ (r+1) \mathcal{P}^{(s)} \right]^{-r/(r+1)}$. 

The Poisson brackets of the charges work out to be \cite{Grumiller-2020}
\begin{equation}\label{eq:gen-charge-algebra}
\begin{aligned}
\{ \mathcal{P}^{(s)}(x), \mathcal{P}^{(s)}(y) \} &= 0~,\\
\{\mathcal{J}_A(x), \mathcal{P}^{(s)}(y)\} &= \left( r\mathcal{P}^{(s)}(y){\partial_{x^A}} - \mathcal{P}^{(s)}(x) {\partial_{y^A}} \right)\delta^2(x-y)~, \\
\{ \mathcal{J}_A(x), \mathcal{J}_B(y) \} &= \left( \mathcal{J}_A(y) {\partial_{x^B}} - \mathcal{J}^{B}(x) 
    {\partial_{y^A}} \right)\delta^2(x-y)~.
\end{aligned}
\end{equation}
This algebra is the semi-direct sum of diffeomorphisms at the spacelike section of the horizon generated by $\mathcal{J}_A$, and an Abelian algebra generated by $\mathcal{P}^{(s)}$. As before, introducing complex coordinates on the sphere, \cite{Grumiller-2020} expanded supertranslations and superrotations in Laurent modes. The corresponding generators can be defined as
\begin{equation}\label{eq:generators}
\begin{gathered}
\mathcal{P}^{(s)}_{mn} = \int \d\zeta \d\Bar{\zeta}~ \zeta^m \Bar{\zeta}^n \mathcal{P}^{(s)}~, \\
\mathcal{J}_m = -\int \d\zeta \d\Bar{\zeta}~ \zeta^{m+1} \mathcal{J}~, \qquad \text{and} \qquad \bar{\mathcal{J}}_m = -\int \d\zeta \d\Bar{\zeta}~ \bar{\zeta}^{m+1} \bar{\mathcal{J}}~,
\end{gathered}
\end{equation}
whose algebra works out to be \cite{Grumiller-2020}
\begin{equation}\label{eq:4d-algebra}
\begin{gathered}
\{\mathcal{J}_k, \mathcal{P}^{(s)}_{(m,n)}\} = \left( \frac{s}{2}(k+1)-m\right) \mathcal{P}^{(s)}_{(k+m,n)}~, \qquad
\{\bar{\mathcal{J}}_k, \mathcal{P}^{(s)}_{(m,n)}\} = \left(\frac{s}{2}(k+1)-n\right)\, \mathcal{P}^{(s)}_{(m,k+n)}  \\ 
\{\mathcal{J}_n, \mathcal{J}_m\} = (n-m)\,\mathcal{J}_{n+m}~, \qquad 
\{\bar{\mathcal{J}}_n,  \bar{\mathcal{J}}_m\} = (n-m)\,\bar{ 
\mathcal{J}}_{n+m}~, \\
\{\mathcal{J}_n,\,\bar{\mathcal{J}}_m\} = 
\{\mathcal{P}^{(s)}_{(m,n)},\,\mathcal{P}^{(s)}_{(m',n')}\} = 0~. 
\end{gathered}
\end{equation}
The third and the fourth brackets --- $\{\mathcal{J}_n, \mathcal{J}_m\}$ and $\{\bar{\mathcal{J}}_n, \bar{\mathcal{J}}_m\}$ --- denote two copies of the Witt algebra. This is the spin-$s$ BMS$_d$ algebra. For $s=0$, this is the algebra for constant surface gravity in \cite{Donnay:2015, Donnay:2016}, which is the Poisson bracket counterpart of \eqref{WittAbelian}. 
The different BMS groups and their relations are shown in Figure \ref{fig:DifferentBMS}.

We would like to briefly note the case of three dimensions since it will be important subsequently. The black hole horizon is then a circle with $A=\phi$ in \eqref{eq:general-equation}.
By expanding the supertranslations and superrotations in Fourier modes, the generators can be defined as
\begin{equation}
\mathcal{P}^{(s)}_n = \frac{1}{2\pi}\,\oint \d\phi\,\mathcal{P}^{(s)}\,e^{\im n\phi}, \qquad 
\mathcal{J}_n =  \frac{1}{2\pi}\,\oint \d\phi\,\mathcal{J}\,e^{\im n\phi}~.
\end{equation}
One gets spin-$s$ BMS$_3$ algebra \cite{Barnich:2006, Grumiller-2020}:
\begin{equation}
\begin{aligned}
        \im\{\mathcal{P}^{(s)}_n,\,\mathcal{P}^{(s)}_m\} &= 0 \\
    \im\{\mathcal{J}_n,\,\mathcal{P}^{(s)}_m\} &= (sn-m)\,\mathcal{P}_{n+m}^{(s)} \\
    \im\{\mathcal{J}_n,\,\mathcal{J}_m\} &= (n-m)\,\mathcal{J}_{n+m}~.
\end{aligned}
\end{equation}

\medskip\noindent\textit{Boundary Conditions for Heisenberg-like Algebra:}  Algebra depends on the choice of slicing used in the solution space \cite{Grumiller-Jabbari-Book}. Interchanging the roles of charges $(\mathcal{P}, \mathcal{J}_A)$ and vectors $(T, \Phi^A)$ in the solution space
\begin{equation}
T = T_H - \Phi^A_H\mathcal{J}_A\mathcal{P}^{-1} \qquad \text{and} \qquad \Phi^A = \Phi^A_H\mathcal{P}^{-1}~.
\end{equation}
the charge variation can also be integrated \cite{Grumiller-2020}. Define $\mathcal{J}^H_A = \mathcal{J}_A\mathcal{P}^{-1}$. Then the charges \eqref{eq:integrated-charge} take the form
\begin{equation}
Q[T_H, \Phi^A_H] = \int \d^2x \left[ T_H\mathcal{P} + \Phi^A_H \mathcal{J}^H_A \right]~,
\end{equation}
with the transformations
\begin{equation}
\delta \mathcal{P} = \delta_A \Phi^A_H \qquad \text{and} \qquad \delta \mathcal{J}^H_A = \delta_A T_H - \Phi^B F_{AB} \mathcal{P}^{-1}~,
\end{equation}
where, $F_{AB} = \partial_A \mathcal{J}^H_B - \partial_B \mathcal{J}^H_A$. The algebra of these charges works out to be
\begin{equation}\label{eq:gen-heisenberg-algebra}
\begin{aligned}
\{ \mathcal{P}(x), \mathcal{P}(y) \} &= 0\\
\{ \mathcal{J}^H_A(x), \mathcal{P}(y) \} &= \partial_{x^A}~\delta(x-y) \\
\{ \mathcal{J}^H_A(x), \mathcal{J}^H_B(y) \} &= \mathcal{P}^{-1}(x) F_{BA}(x) \delta(x-y)~.
\end{aligned}
\end{equation}
When $F_{AB} = 0$, this is exactly Heisenberg algebra \cite{Grumiller-2020}.

\section{Local Axisymmetric Black Holes}\label{sec:local-black-holes}
All explicit examples studied in four dimensions have exclusively considered near-horizon symmetries of black holes with spherical horizon topology, including the works mentioned above. We now consider axisymmetric black holes in vacuum whose horizons could be either spherical or toroidal. 

The general static axisymmetric metric can be written in the following gauge
\begin{equation}\label{eq:general-equation}
ds^2 = -F G \d t^2 + \frac{F}{G} \d\eta^2 + \hat{\Omega}_{AB}\d\phi^A \d\phi^B~,
\end{equation}
where, $\eta$ is the radial coordinate, $\phi^A$ denotes the angles with $A \in \{2,3\}$, and all metric coefficients in \eqref{eq:general-equation} are functions of $\eta$ and $\theta (\equiv \phi^{A=2})$. Let $\eta_0$ be the Killing horizon; $F$ and $G$ are chosen in such a way that $G(\eta_0,\theta) = 0$, $G_{,\eta} (\eta_0,\theta) \neq 0$, and $F(\eta_0,\theta) \neq 0$ (can be zero at the extremities of $\theta$). We will assume that the black hole is non-extremal, meaning that the zeros of $G(\eta_0,\theta)$ are not degenerate.

To study the near-horizon geometry, the appropriate new radial coordinate that vanishes on the boundary works out to be
\begin{equation}\label{eq:NH-rho}
\rho^2 = 4(\eta-\eta_0) \left(\frac{F}{G_{,\eta}}\right)_{\eta=\eta_0}.
\end{equation}
This will ensure the Rindler-like expansion of the near-horizon geometry.
\subsection*{Spherical Horizons}\label{sec:spherical-horizons}
Chandrasekhar obtained all vacuum spherical topology axisymmetric black holes in Schwarzschild-like coordinates \cite{chandrasekharbook, LocalToroidalXanthapolous}:
\begin{equation}\label{sphericallocalmetric}
ds^2 = -\frac{\eta-1}{\eta+1} e^S \d t^2 + \frac{\eta +1}{\eta-1} m^2 e^{\sigma - S} \d\eta^2 + m^2(\eta+1)^2 e^{-S} \left[ e^\sigma \d\theta^2 + \sin^2(\theta) \d\phi^2 \right]
\end{equation}
where $\eta\equiv (r-m)/m$ is a radial coordinate, and $m$ represents the ``mass" of the black hole. For this to be a vacuum solution, the function $S(\eta,\mu)$ must take the form
\begin{equation}
S(\eta,\mu) = \sum^\infty_{n=0} A_n P_n(\eta) P_n(\mu)~,
\end{equation}
where $\mu=\cos\theta$, $A_n$ real constants, $P_n$ Legendre polynomial, and the function $\sigma(\eta,\mu)$ is obtained by a line integral from
\begin{subequations}\label{sigmaequations}
\begin{align}
\frac{\eta^2-\mu^2}{(\eta^2-1)(1-\mu^2)}\sigma_{,\eta} &= \frac{2\eta}{\eta^2-1}S_{,\eta} - \frac{2\mu}{\eta^2-1}S_{,\mu} - \mu S_{,\eta}S_{,\mu} + \frac{\eta}{2(\eta^2-1)}\left[(\eta^2-1)S^2_{,\eta}+(\mu^2-1)S^2_{,\mu}\right]~, \\
\frac{\eta^2-\mu^2}{(\eta^2-1)(1-\mu^2)}\sigma_{,\mu} &= \frac{2\mu}{1-\mu^2}S_{,\eta} + \frac{2\eta}{\eta^2-1}S_{,\mu} + \eta S_{,\eta}S_{,\mu} + \frac{\mu}{2(1-\mu^2)}\left[(\eta^2-1)S^2_{,\eta}+(\mu^2-1)S^2_{,\mu}\right]~.\label{sig_2}
\end{align}
\end{subequations}
The horizon is the hypersurface $\eta=1$, and hence, equation \eqref{sig_2} implies that for the black hole horizon to be regular, the condition
\begin{equation}\label{sph_reg1}
\sigma_{,\mu} = 2S_{,\mu} \quad \text{at} \quad \eta = 1
\end{equation}
must be satisfied. One can rewrite this as $\sigma(1,\mu) = 2S(1,\mu) + 2\alpha$ for some constant $\alpha$. That the horizon is locally flat near the axis of symmetry (equivalently, the horizon is topologically $\mathrm{S}^2$ \cite{LocalToroidalXanthapolous}) puts a constraint on the odd coefficients
\begin{equation}\label{sph_reg2}
\sum^\infty_{n=0} A_{2n+1} = 0~.
\end{equation}
Equations \eqref{sph_reg1} and \eqref{sph_reg2} collectively ensure the horizon is regular and there are no conical singularities on the two poles of the horizon. Note that $S=0$ $(\implies \sigma = 0)$ is the Schwarzschild metric. 

The new near-horizon radial coordinate \eqref{eq:NH-rho} is then
\begin{equation}
\rho^2 = 4(\eta-1) \left( \frac{m e^{\frac{\sigma}{2}}}{G_{,\eta}} \right)_{\eta=1}~,
\end{equation}
where
\begin{equation}
G = \frac{\eta-1}{m(\eta+1)} \e^{S-\frac{1}{2}\sigma}~.
\end{equation}
With this, the metric coefficients fall-off near the horizon as in equation \eqref{eq:Fall-off}
with
\begin{equation}
\begin{aligned}
\kappa &= \frac{{\mathrm e}^{-\alpha}}{4m}~, & R_A &= 0 , & M_A &= \left( 1-\mu^2 \right) \left(S''(1,\mu) - \sigma''(1,\mu)\right)\\ 
\Omega_{\theta\theta} &= 4m^2 e^{2\alpha + S(1,\mu)}~, & \Omega_{\phi\phi} &= 4m^2 \sin^2\theta e^{-S(1,\mu)}~, & \Omega_{\theta\varphi} &= 0~.
\end{aligned}
\end{equation}
Thus, the Killing vectors that preserve the fall-off conditions are just the same as given by \eqref{eq:Killing-Vectors}. The functions $\mathcal{P}$ and $\mathcal{J}_A$ from \eqref{eq:p-and-j} are given by
\begin{equation}
\mathcal{P} = \frac{\e^{\alpha}m^2 \sin\theta}{2\pi} \qquad \text{and} \qquad \mathcal{J}_A = 0~.
\end{equation}
Note that since $R_A = 0$, $\mathcal{J}_A$ will also be zero. However, since $\delta R_A$ need not be zero, the integrated charge associated with $R_A$ will not vanish. We will be using $R_A$ and $\mathcal{J}_A$ to denote both the metric-related terms and the integrated terms ($R_A$ and $\mathcal{J}_A$ below refer to the integrated terms). Since the horizon is topologically a sphere, we can use the methods in Section \ref{sec:gen-treatment} to get spin-$s$ BMS$_4$ algebra.

Using equation \eqref{eq:Hamiltonian}, the Hamiltonian works out to be (after substituting $\mathcal{N}^{(s)}$ and $\mathcal{P}^{(s)}$ in terms of $\mathcal{P}$)
\begin{equation}\label{eq:sph-hamil-firstlaw}
H = \int \d^2 x \mathcal{N}^{(s)}\mathcal{P}^{(s)} = \int \d^2x \kappa \mathcal{P} = \frac{\kappa A}{8\pi} = \frac{m}{2}~.
\end{equation}
Note that this equation produces the correct first law of black hole thermodynamics: $\delta H = \kappa/(2\pi)~ \delta (A/4)$, a point we would return to in section \ref{sec:thermodynamics}.

The algebra obtained through the Heisenberg-like boundary condition for local spherical black holes would not differ either from the algebra of spherical black holes considered in \cite{Grumiller-2020}. This is because the topology of the horizon determines the algebra, as mentioned earlier.

\subsection*{Toroidal Horizons}\label{sec:toroidal}
Following Chandrasekhar, Xanthopoulos worked out all axially symmetric toroidal local black holes in vacuum in similar coordinates \cite{LocalToroidalXanthapolous} 
\begin{equation}
ds^2 = -\frac{1}{4}(\eta^2-1)(1-\mu^2) e^{-S} \d t^2 + 4m^2 e^{S}\d\varphi^2 + 4m^2(\eta^2-\mu^2)e^{\sigma-S}\left[\frac{\d\eta^2}{\eta^2-1}+\frac{\d\mu^2}{1-\mu^2}\right]
\end{equation}
where $\eta$ is the radial coordinate, $m$ represents the ``total mass of the background black hole", and $\mu = \cos\theta$. The function $S(\eta,\mu)$ is given by
\begin{equation}
S(\eta,\mu) = \sum^\infty_{n=0} B_n P_n(\eta) P_n(\mu)~,
\end{equation}
where $B_n$'s are constants and $P_n$ the Legendre polynomials. The function $\sigma(\eta,\mu)$ is obtained from $S$ (up to an additive constant) by solving
\begin{subequations}\label{sigmaequationstoro}
\begin{align}
\sigma_{,\eta} &= -\frac{\eta(\mu^2-1)}{2(\eta^2-\mu^2)}\left[ (\eta^2-1)S_{,\eta}^2 + (\mu^2-1)S_{,\mu}^2 \right] + \frac{\mu(\eta^2-1)(\mu^2-1)}{\eta^2-\mu^2}S_{,\eta}S_{,\mu}\\[10pt]
\sigma_{,\mu} &= \frac{\mu(\eta^2-1)}{2(\eta^2-\mu^2)}\left[ (\eta^2-1)S_{,\eta}^2 + (\mu^2-1)S_{,\mu}^2 \right] - \frac{\eta(\eta^2-1)(\mu^2-1)}{\eta^2-\mu^2}S_{,\eta}S_{,\mu}~
\end{align}
\end{subequations}
These equations are different from the spherical-topology case (i.e., equation \eqref{sigmaequations}). In particular, the right-hand side here does not contain any linear derivative of $S$. As before, $\sigma$ is constant on the horizon $\eta=1$, which we will call $2\alpha$ again. The conditions for horizon-regularity are different \cite{LocalToroidalXanthapolous} 
\begin{equation}\label{regtoroidal}
\sum_{n=0}^\infty B_{2n+1} = 0~, \qquad  \sum_{n=0}^\infty B_{2n}P_{2n}^{\prime}(1) = 0~, \qquad  \sum_{n=0}^\infty B_{2n+1}P_{2n+1}^{\prime\prime}(1) = 0~,
\end{equation}
where prime denotes differentiation with respect to $\mu(\equiv \cos\theta)$.

The near-horizon radial coordinate \eqref{eq:NH-rho} works out to be
\begin{equation}
\rho^2 = 4(\eta-1) \left( m e^{-S+\frac{1}{2}\sigma} \frac{\sqrt{(\eta^2-\mu^2)(1-\mu^2)}}{G_{,\eta}} \right)_{\eta=1}~,
\end{equation}
with
\begin{equation}
G = \frac{\left(\eta^{2}-1\right) {\mathrm e}^{-\frac{\sigma \left(\eta ,\mu \right)}{2}}}{4 m}  \sqrt{\frac{1-\mu^{2}}{\eta^{2}-\mu^{2}}}~.
\end{equation}  
With 
\begin{equation}
\begin{aligned}
\kappa &= \frac{{\mathrm e}^{-\alpha}}{4m}~, & R_A &= 0~, & M_\theta &= S''(1,\mu) - \frac{1}{2} \sigma''(1,\mu) + \frac{2\mu}{1-\mu^2} \\ 
\Omega_{\mu\mu} &= 4m^2 e^{2\alpha -S(1,\mu)}~, & \Omega_{\phi\phi} &= 4m^2 e^{S(1,\mu)}~, & \Omega_{\mu\varphi} &= 0,
\end{aligned}
\end{equation}
one gets the desired fall-off conditions where the specific functions are as follows. Thus the Killing vectors preserving the fall-offs are
\begin{equation}
\xi^t = \frac{1}{\kappa} T(t, \phi^A) + \mathcal{O}(\rho)~, \qquad \xi^\rho = \mathcal{O}(\rho^2)~, \qquad \xi^A = \Phi^A(\phi^B) + \mathcal{O}(\rho^2)~.
\end{equation}
The three dynamical variables are 
\begin{equation}
\mathcal{P} = \frac{\e^{\alpha}m^2}{2\pi} \qquad \text{and} \qquad \mathcal{J}_A = 0~.
\end{equation}
Note that we do not have the $\sin{\theta}$ factor in $\mathcal{P}$ here as in the spherical case. This is because we are essentially dealing with two flat directions. As mentioned earlier, $\mathcal{J}_A = 0$ does not mean that the charge associated with it will be zero.  The Hamiltonian is given by
\begin{equation}
H = \int \d^2 x \mathcal{N}^{(s)}\mathcal{P}^{(s)} = \int \d^2x \kappa\mathcal{P} = \frac{\kappa A}{8\pi} = \frac{m}{2}~,
\end{equation}
which is the same as in the case of spherical horizons.

\subsection*{Algebras}
\subsubsection*{Killing Vector Algebra}
We first work out the closed form of $[\xi(T_1,\Phi_1^A), \xi(T_2, \Phi_2^A)]$, i.e., the Lie bracket algebra of the asymptotic Killing vector fields starting from the equation \eqref{eq:function-expansion-spherical}. Since now the arbitrary functions $T$ and $\Phi^A$ are functions on a flat torus, we can expand them in Fourier modes as follows:
\begin{equation}
T = \sum_{\alpha, \beta \in \mathbb{Z}} \tau_{\alpha\beta} \e^{\im (\alpha\theta_1 + \beta\theta_2)}~, \qquad \Phi^A = \sum_{\alpha, \beta \in \mathbb{Z}} y^A_{\alpha\beta} \e^{\im (\alpha\theta_1 + \beta\theta_2)}~
\end{equation}
where $0\le \theta_1,\theta_2 \le 2\pi$. A basis for the vector space is then given by
\begin{equation}
\mathcal{T}_{mn} = \xi\left( \e^{\im(m\theta_1 + n\theta_2)}, 0, 0 \right)~, \qquad  \mathcal{Y}^2_{mn} =  \xi\left(0, \e^{\im(m\theta_1 + n\theta_2)}, 0 \right)~, \qquad \mathcal{Y}^3_{mn} = \xi\left(0, 0, \e^{\im(m\theta_1 + n\theta_2)}\right)~.
\end{equation}
These give the following nonzero commutation relations 
\begin{equation}\label{eq:Toroidal-Lie}
\begin{aligned}
\relax
\im [\mathcal{Y}^2_{pq}, \mathcal{Y}^2_{mn}] &= (p-m) \mathcal{Y}^2_{p+m, q+n}~, & 
\im [\mathcal{Y}^3_{pq}, \mathcal{Y}^3_{mn}] &= (q-n) \mathcal{Y}^3_{p+m, q+n}~, \\ \im [\mathcal{Y}^2_{pq}, \mathcal{Y}^3_{mn}] &= q\mathcal{Y}^2_{p+m, q+n} - m\mathcal{Y}^3_{p+m, q+n}~, &
\im [\mathcal{T}_{pq}, \mathcal{Y}^2_{mn}] &= p \mathcal{T}_{p+m, q+n}~, \\
\im [\mathcal{T}_{pq}, \mathcal{Y}^3_{mn}] &= q \mathcal{T}_{p+m, q+n}~.
\end{aligned}
\end{equation}
Thus, the algebra consists of two \emph{non-commuting} copies of the Witt algebra generated by $\mathcal{Y}^2_{mn}$ and $\mathcal{Y}^3_{mn}$, and an Abelian current algebra generated by $\mathcal{T}_{pq}$.

\subsubsection*{Charge Algebras} 
\textbf{BMS-like Boundary Conditions}: 
We now find the algebra of the charges starting from equation \eqref{eq:4d-algebra}. Again, we expand the generators in Fourier modes as follows:
\begin{equation}\label{eq:Toroidal-Fourier}
\mathcal{P}^{(s)}_{mn} = \frac{1}{4\pi^2} \int \d\theta_1 \d\theta_2 \mathcal{P}^{(s)} \e^{\im m\theta_1} \e^{\im n\theta_2} \qquad \text{and} \qquad \mathcal{J}^A_{mn} = \frac{1}{4\pi^2} \int \d\theta_1 \d\theta_2 \mathcal{J}_A \e^{\im m \theta_1} \e^{\im n\theta_2}
\end{equation}
This leads to the following algebra (details in Appendix \ref{appendix:toroidal-algebra}):
\begin{equation}\label{eq:gen-toroidal-algebra}
\begin{aligned}
    \im\{\mathcal{J}^A_{kl},~\mathcal{P}^{(s)}_{mn}\} &= \left[\left(\frac{s}{2} k-m\right)\delta_{A2} + \left(\frac{s}{2} l-n \right)\delta_{A3} \right]~\mathcal{P}_{k+m~l+n}^{(s)} \\
    \im\{\mathcal{P}^{(s)}_{kl},~\mathcal{P}^{(s)}_{mn}\} &= 0 \\
    \im\{\mathcal{J}^A_{kl},~\mathcal{J}^B_{mn}\} &= \left[ k\delta_{A2} + l\delta_{A3}\right]~\mathcal{J}^B_{k+m~l+n} - \left[ m\delta_{B2} + n\delta_{B3}\right]~\mathcal{J}^A_{k+m~l+n}
\end{aligned}
\end{equation}
where, $\delta_{Ai}$ here is the Kronecker delta. Written explicitly by considering $\mathcal{J}^2$ and $\mathcal{J}^3$ separately, the algebra is given by
\begin{equation}\label{eq:toroidal-algebra}
\begin{aligned}
    \im\{\mathcal{J}^2_{kl},~\mathcal{P}^{(s)}_{mn}\} &= \left(\frac{s}{2}k-m\right)~\mathcal{P}_{k+m~l+n}^{(s)}  &\qquad     \im\{\mathcal{J}^3_{kl},~\mathcal{P}^{(s)}_{mn}\} &= \left(\frac{s}{2}l-n\right)~\mathcal{P}_{k+m~l+n}^{(s)} \\
    \im\{\mathcal{P}^{(s)}_{kl},~\mathcal{P}^{(s)}_{mn}\} &= 0 &\qquad     \im\{\mathcal{P}^{(s)}_{kl},~\mathcal{P}^{(s)}_{mn}\} &= 0 \\
    \im\{\mathcal{J}^2_{kl},~\mathcal{J}^2_{mn}\} &= (k-m)~\mathcal{J}^2_{k+m~l+n} &\qquad     \im\{\mathcal{J}^3_{kl},~\mathcal{J}^3_{mn}\} &= (l-n)~\mathcal{J}^3_{k+m~l+n}~,\\
    \im\{\mathcal{J}^2_{kl},~\mathcal{J}^3_{mn}\} &= k~\mathcal{J}^3_{k+m~l+n} -  n~\mathcal{J}^2_{k+m~l+n} &= -  \im\{\mathcal{J}^3_{mn},~\mathcal{J}^2_{kl}\}~.
\end{aligned}
\end{equation}
To summarize, the algebra is the semi-direct sum of diffeomorphisms at the space-like section of the horizon (that is, $\mathrm{S}\times \mathrm{S}$ for the toroidal) generated by $\mathcal{J}_A$, and an Abelian current generated by $\mathcal{P}^{(s)}$. Note that each of the pairs $\mathcal{P}^{(s)}_{kl}$, $\mathcal{J}^2_{mn}$ and $\mathcal{P}^{(s)}_{kl}$, $\mathcal{J}^3_{mn}$ forms a BMS$_3$ sub-algebra \cite{Grumiller-2020}. This can easily be verified by mapping
\begin{equation}
\begin{aligned}
    \text{for} \quad &\mathcal{P}^{(s)}_{kl}, \mathcal{J}^2_{mn}: & & & \text{for} \quad &\mathcal{P}^{(s)}_{kl}, \mathcal{J}^3_{mn}:\\
    \mathcal{P}^{(s)}_{mn} &\mapsto \mathcal{P}^{(s/2)}_{m} & & \text{and} & \mathcal{P}^{(s)}_{mn} &\mapsto \mathcal{P}^{(s/2)}_{n}~~~~.\\
    \mathcal{J}^2_{kl} &\mapsto \mathcal{J}_{k} & & & \mathcal{J}^3_{kl} &\mapsto \mathcal{J}_{l}
\end{aligned}   
\end{equation}

\noindent\textbf{Heisenberg-like Boundary Condition:} 
As before, the generators can be expanded as follows
\begin{equation}
\mathcal{P}_{mn} = \frac{1}{4\pi^2} \int \d\theta_1 \d\theta_2 \mathcal{P} \e^{\im m\theta_1} \e^{\im n\theta_2} \qquad \text{and} \qquad \mathcal{J}^A_{mn} = \frac{1}{4\pi^2} \int \d\theta_1 \d\theta_2 \mathcal{J}^H_A \e^{\im m\theta_1} \e^{\im n\theta_2}
\end{equation}
Using \eqref{eq:gen-heisenberg-algebra}, the Heisenberg-like algebra of the charges in this case work out to be
\begin{equation}\label{eq:toroidal-algebra-Heisenberg}
\begin{aligned}
    \im\{\mathcal{J}^2_{kl},~\mathcal{P}_{mn}\} &= 0  &\qquad     \im\{\mathcal{J}^3_{kl},~\mathcal{P}_{mn}\} &= 0 \\
    \im\{\mathcal{P}_{kl},~\mathcal{P}_{mn}\} &= 0 &\qquad     \im\{\mathcal{P}_{kl},~\mathcal{P}_{mn}\} &= 0 \\
    \im\{\mathcal{J}^2_{kl},~\mathcal{J}^2_{mn}\} &= 0 &\qquad     \im\{\mathcal{J}^3_{kl},~\mathcal{J}^3_{mn}\} &= 0~,
\end{aligned}
\end{equation}
\begin{equation}\label{eq:toroidal-algebra-Heisenberg-NC}
    \{\mathcal{J}^2_{kl},~\mathcal{J}^3_{mn}\} = \int \d\theta_1 \d\theta_2 \mathcal{P}^{-1}\left( \partial_{\theta_2} \mathcal{J}^H_2 - \partial_{\theta_1} \mathcal{J}^H_3 \right) \e^{\im\left[ (k+m)\theta_1 + (l+n)\theta_2 \right]} = -  \{\mathcal{J}^3_{mn},~\mathcal{J}^2_{kl}\}~.
\end{equation}
The last bracket \eqref{eq:toroidal-algebra-Heisenberg-NC} cannot be reduced further because of the nonlinearity of the expression, except perhaps for rather special choices of $\mathcal{P}$. The supertranslations $\mathcal{P}_{kl}$ commute with all other generators, same as in the spherical case. The two generators of the superrotations $\mathcal{J}^A_{mn}$, on the other hand, have a nonzero commutation between them.

\section{Thermodynamics}\label{sec:thermodynamics}
We now take a closer look at the thermodynamics of general $s$ formalism. As we have seen in section \ref{sec:gen-treatment}, the total Hamiltonian takes the form
\begin{equation} \label{eq:hamiltonian}
H = \int \d^{d-2}x\, \mathcal{N}^{(s)} \mathcal{P}^{(s)}~.
\end{equation}
Assuming constant $\kappa$, and using $\mathcal{N}^{(s)} = \kappa \left[ (r+1) \mathcal{P}^{(s)} \right]^{-r/(r+1)}$ and equation \eqref{eq:t-p-r}, to rewrite the Hamiltonian in terms of $\mathcal{P}$, one obtains (this is essentially what one obtains as the Hamiltonian in a co-rotating frame)
\begin{equation}
H = \int \d^{d-2}x \kappa \mathcal{P} = \frac{\kappa A}{8\pi}~,
\end{equation}
whose variation gives
\begin{equation}
\delta H = \frac{\kappa}{2\pi} \delta\left(\frac{A}{4}\right) = \beta^{-1} \delta S~,
\end{equation}
where $A$ is the area of the horizon, $S$ is the black hole entropy, and $\beta$ is the Euclidean time period with $T=\beta^{-1}$ the Hawking temperature of the hole.
The introduction of a spin parameter should not, in principle, affect the first law. However, as we will see below, that is not the case.
Rewriting the entropy as
\begin{equation}\label{eq:entropy}
S = \frac{A}{4} = 2\pi \int d^{d-2}x \frac{\sqrt{\Omega}}{8\pi} = 2\pi \int \d^{d-2}x \mathcal{P}~,
\end{equation}
and expressing it in terms of $\mathcal{P}^{(s)}$ using $\mathcal{P} = \left(r+1\right)^{\frac{1}{r+1}} \left(\mathcal{P}^{(s)}\right)^{\frac{1}{r+1}}$, one obtains
\begin{equation}
S = 2\pi \left( r+1 \right)^{\frac{1}{r+1}} \int \d^{d-2}x \left(\mathcal{P}^{(s)}\right)^{\frac{1}{r+1}}~,
\end{equation}
where, as before, $r=s/(d-2)$.

One can likewise write the Hamiltonian \eqref{eq:hamiltonian} in terms of $\mathcal{P}^{(s)}$ and $\kappa$:
\begin{equation}
H = \int \d^{d-2}x \kappa \left(r+1\right)^{-\frac{r}{r+1}} \left(\mathcal{P}^{(s)}\right)^{\frac{1}{r+1}}~.
\end{equation}
From the above two equations, for constant surface gravity $\kappa$, one obtains
\begin{equation}\label{thetwovariations}
\delta H = \kappa \left(r+1\right)^{-\frac{r}{r+1}} \int \d^{d-2}x \delta \left(\mathcal{P}^{(s)}\right)^{\frac{1}{r+1}} \qquad \text{and} \qquad \delta S = 2\pi \left( r+1 \right)^{\frac{1}{r+1}} \int \d^{d-2}x \delta \left(\mathcal{P}^{(s)}\right)^{\frac{1}{r+1}}~,
\end{equation}
which gives
\begin{equation}
\delta H = \frac{\kappa}{2\pi (r+1)} \delta S~.
\end{equation}
Thus, for any nonzero value of $s$ we do not obtain the usual first law\footnote{The difference in the prefactors in the two variations in \eqref{thetwovariations} seems to have been overlooked in \cite{Grumiller-2020}, which may have contributed to the conclusion that the first law holds.}. One can try redefining the Hawking temperature as 
\begin{equation}
T=\frac{\kappa}{2\pi} \rightarrow{\frac{\kappa}{2\pi (r+1)}}~.    
\end{equation} 
However, this freedom does not exist since $\kappa$ is the hole's surface gravity determined from the metric's expansion, equation \eqref{eq:Fall-off}. Recall that in Euclidean quantum gravity, $\kappa$ determines the periodicity of the $S^1$ fiber. Rescaling it (equivalently, rescaling the time coordinate) would result in a codimension-two Euclidean space with a conical singularity, and such geometries cannot be included in the path integral.\footnote{This conical singularity is different from the conical singularities on the horizon discussed earlier.} Note that the integral \eqref{eq:entropy} for black hole entropy does not contain $s$, so the second law of thermodynamics holds. This also means $\delta S$ cannot accommodate the $r+1$ factor.

The difference essentially arises from the extra $r+1$ factor in equation \eqref{eq:integrated-chargewithr}
\begin{equation*}
Q = \int \d^{d-2}x \left[ T \frac{\mathcal{P}}{r+1} + \Phi^A \mathcal{J}_A \right]~ \tag{\ref{eq:integrated-chargewithr}}
\end{equation*}
compared to equation (\ref{eq:integrated-charge})\footnote{This can be written as $Q = \frac{1}{16\pi} \int \d^{d-2}x  \left[ 2T\sqrt{\Omega} + 2 \Phi^A \frac{\sqrt{\Omega}}{{\kappa}} \left( \partial_t M_A - R_A \right) \right]$, and is similar to equation (2.32) in \cite{Donnay:2016}, and equation (14) in \cite{Donnay:2019}, expect for slight notational differences.}
\begin{equation}
Q = \int \d^{d-2}x \left[ T \mathcal{P} + \Phi^A \mathcal{J}_A \right]~. \tag{\ref{eq:integrated-charge}}
\end{equation}
Both of these integrated charges are consistent with the charge variation equation \eqref{eq:charge-variation}. The $r+1$ factor in \eqref{eq:integrated-chargewithr} is essential in obtaining the BMS$_d$ algebra. 

Instead of rescaling the temperature, one can rescale the Hamiltonian factor to preserve the first law. This can be done by redefining the chemical potential $\mathcal{N}^{(s)}$ as $\kappa \left(r+1\right)^{1/(r+1)} \left(\mathcal{P}^{(s)} \right)^{-r/(r+1)}$. However, in this case, the total Hamiltonian will no longer generate unit-time translations. The other problem is that this would introduce the spin parameter in the Hamiltonian (even when written in terms of $\mathcal{P}$), meaning that the $s\neq 0$ soft-hair terms would contribute to the energy. Thus,  we find no immediate way to reconcile the first law for nonzero $s$ while maintaining constant surface gravity. It may still be possible to reconcile it with constant $\kappa$ under certain special thermodynamic configurations. For example, as a purely speculative idea, the rescaled temperature could represent the local temperature of an isothermal cavity in a heat bath (see, for example, \cite{York-Brown:1992, Akbar-Gibbons-2003, Akbar:2010}). Whether such an interpretation is possible remains unclear and is left for future investigation.

In summary, there is no issue with the first law when the Hamiltonian is kept free of the spin parameter. However, once the Hamiltonian is expressed in terms of the spin parameter and thermal equilibrium is assumed for the black hole, the first law is violated. This raises the question of how the boundary conditions used to obtain the BMS$_d$ algebra and the spin parameter should be interpreted. Unable to find an immediate resolution, we are led to believe that the BMS algebra boundary condition applies to dynamical black holes (with varying $\kappa$), meaning the black hole is not in equilibrium with the thermal bath with which it interacts. It remains unclear how to formulate this interaction in terms of a thermodynamic configuration.

\section{Conclusion}
Near-horizon symmetries can be approached in several ways, subject to different boundary conditions, all giving supertranslations and superrotations but generally differing in their algebras, {depending on the spacetime dimension and the topology of the black holes}. For isolated holes, the algebra was found to be two copies of Witt and an abelian current \cite{Donnay:2015, Donnay:2016, Donnay:2019}. By formulating a suitable boundary condition in the ADM formalism ingeniously parameterized by a parameter, it was subsequently found that one can obtain the BMS algebra.
While the BMS group is necessarily the symmetry group on $\mathscr{I}$ (which is topologically $\mathrm{S}^2 \times \mathbb{R}$) for asymptotically flat spacetimes, there is no a priori reason for this to be the preferred group for near-horizon symmetries of black holes even when the topology of the horizon is $\mathrm{S}^2 \times \mathbb{R}$. Thus, this result came as a welcome surprise\footnote{As long as the conserved charges on the horizon outnumber the charges at infinity, we have a satisfactory scattering process in the sense of \cite{Hawking:2016}.}.

All explicit examples previously studied in connection with near-horizon symmetries (in four dimensions) have strictly (geometrical) spherical horizons, with the only exceptions being the C-metrics \cite{Anabalon2021}, which have axially symmetric horizons that are topological two-spheres. The only study of non-spherical topology in near-horizon symmetries to date examines black holes with an $\mathrm{S}^1 \times \mathrm{S}^2$ horizon topology in five dimensions \cite{Giribet2023}. In this paper, we have examined static black holes in four-dimensional vacuum general relativity that are topologically spherical and toroidal but geometrically axially symmetric. None of these solutions, while solving the vacuum Einstein equations, are asymptotically flat. These solutions are regular everywhere, and their horizons are non-singular, unlike the vacuum C-metric, and together they constitute the complete set of axisymmetric static black holes that solve the vacuum Einstein equations. We have found that the symmetry algebra for the toroidal black holes changes to being two copies of the spin-$s$ BMS$_3$ from spin-$s$ BMS$_4$  for spherical topology black holes. For $s=0$, this gives two noncommuting Witt algebras for the toroidal holes.

{ Intuitively, one may expect that algebra, being local, would not be affected by the topology of the horizon. However, as we have seen in this paper, the infinite-dimensional near-horizon symmetry group involves arbitrary functions defined on the horizon, and the space of allowed functions on a torus differs from that on a sphere. This causes the generators to depend on the topology of the horizon, resulting in different algebras. It would be interesting to investigate how topology affects near-horizon symmetries in a more general setting. For example, one could study higher-genus black holes in four (or higher) dimensions in the presence of a negative cosmological constant \cite{Brill_1997}. This is left for future work.}

Returning to the general formulation, we reconsider thermodynamics for the spin-$s$ BMS$_d$. For the $s \ne 0$ BMS algebra, we find it at odds with the first law of thermodynamics for constant $\kappa$ (temperature), i.e., for isolated black holes. Redefining the temperature (or the Hamiltonian) by the extra factor that arises does not resolve the issue. We are thus led to conclude that the BMS boundary condition fundamentally applies to black holes that are thermodynamically dynamic. However, in the limit $s = 0$, the first law holds for isolated black holes. In this case, the near-horizon symmetry algebra yields two copies of the Witt algebra (commuting for spherical and noncommuting for toroidal holes) and an Abelian current.

\section*{Acknowledgements}
We thank Carlos Arreche, Charlie Brewer, Gaston Giribet, Sri Rama Chandra Kushtagi, and Mohammad Mehdi Sheikh-Jabbari for their helpful comments.

\appendix

\section{Toroidal Algebra}\label{appendix:toroidal-algebra}
Here we present detailed calculations for one of the Lie brackets for the toroidal black holes \eqref{eq:Toroidal-Lie}. Consider the Lie bracket $[\mathcal{Y}^2_{pq}, \mathcal{Y}^3_{mn}]$. Making use of equation \eqref{eq:LieBracket-detail}, we immediately see $T_{12} = 0$. The other two relations work out to be
\begin{equation}
\begin{aligned}
\Phi^\theta_{12} &= \Phi^B_1 \partial_B \Phi^\theta_2 - \Phi^B_2 \partial_B \Phi^\theta_1 =  \cancelto{0}{\e^{\im (p\theta + q\phi)}\partial_\theta \Phi^\theta_2} - \e^{\im(m\theta + n\phi)} \partial_\phi \e^{\im(p\theta + q\phi)}~, \\
\Phi^\phi_{12} &= \Phi^B_1 \partial_B \Phi^\theta_1 - \Phi^B_2 \partial_B \Phi^\phi_1 =  \e^{\im (p\theta + q\phi)}\partial_\theta \e^{\im(m\theta + n\phi)} - \cancelto{0}{\e^{\im(m\theta + n\phi)} \partial_\phi \Phi^\theta_1}~. \\
\end{aligned}
\end{equation}
Thus, the bracket computes to 
\begin{equation}
[\mathcal{Y}^2_{pq}, \mathcal{Y}^3_{mn}] = -\im q \e^{\im \left[ (p+m)\theta + (q+n)\phi \right]} \partial_\theta + \im m \e^{\im \left[ (p+m)\theta + (q+n)\phi \right]} \partial_\phi = -\im q \mathcal{Y}^2_{p+m, q+n} + \im m \mathcal{Y}^3_{p+m, q+n}~.
\end{equation}
Similarly, all other brackets can be computed.

Now the detailed calculation for one of the toroidal charge algebras \eqref{eq:toroidal-algebra}. Consider the bracket $\{\mathcal{J}^A_{kl},~\mathcal{J}^B_{mn}\}$. Using the Fourier expansion \eqref{eq:Toroidal-Fourier}, this can be written as
\begin{equation}
\{\mathcal{J}^A_{kl},~\mathcal{J}^B_{mn}\} = \frac{1}{4\pi} \int \d^2x \{\mathcal{J}^A(x_2,x_3), \mathcal{J}^B(y_2,y_3)\} \e^{\im\left( k x_2 + l x_3 + m y_2 + n y_3 \right)}~.
\end{equation}
We know from \eqref{eq:gen-charge-algebra} that
\begin{multline}
\frac{1}{4\pi} \int \d^2x \{\mathcal{J}^A(x_2,x_3), \mathcal{J}^B(y_2,y_3)\} \e^{\im\left( k x_2 + l x_3 + m y_2 + n y_3 \right)} \\ = \frac{1}{4\pi} \int \d^2x \left( \mathcal{J}^A \frac{\partial}{\partial y_B} - \mathcal{J}^A \frac{\partial}{\partial x_B} \right) \e^{\im\left( k x_2 + l x_3 + m y_2 + n y_3 \right)} \delta(\mathbf{x}-\mathbf{y})
\end{multline}
Here one needs to be careful about which derivative acts where. The above expression computes to
\begin{multline}
\frac{1}{4\pi} \int \d^2x \left( \mathcal{J}^A \frac{\partial}{\partial y_B} - \mathcal{J}^B \frac{\partial}{\partial x_A} \right) \e^{\im\left( k x_2 + l x_3 + m y_2 + n y_3 \right)} \delta(x_2-y_2) \delta(x_3-y_3) \\ 
= \frac{1}{4\pi} \int \d^2x \left( \im \left[ m\delta_{B2} + n\delta_{B3} \right] \mathcal{J}^A - \im \left[ k\delta_{A2} + l\delta_{A3} \right] \mathcal{J}^B \right) \e^{\im\left( k x_2 + l x_3 + m y_2 + n y_3 \right)} \delta(x_2-y_2) \delta(x_3-y_3) 
\end{multline}
which gives
\begin{equation}
\{\mathcal{J}^A_{kl},~\mathcal{J}^B_{mn}\} = \im \left[ m\delta_{B2} + n\delta_{B3} \right] \mathcal{J}^A_{k+m~l+n} - \im \left[ k\delta_{A2} + l\delta_{A3} \right] \mathcal{J}^B_{k+m~l+n}~.
\end{equation}
Written explicitly, we have
\begin{subequations}
\begin{equation}
    \im\{\mathcal{J}^2_{kl},~\mathcal{J}^2_{mn}\} = (k-m)~\mathcal{J}^2_{k+m~l+n}~, \qquad     \im\{\mathcal{J}^3_{kl},~\mathcal{J}^3_{mn}\} = (l-n)~\mathcal{J}^3_{k+m~l+n}~,
\end{equation}
and
\begin{equation}
    \im\{\mathcal{J}^2_{kl},~\mathcal{J}^3_{mn}\} = k~\mathcal{J}^3_{k+m~l+n} - n~\mathcal{J}^2_{k+m~l+n} = -  \im\{\mathcal{J}^3_{mn},~\mathcal{J}^2_{kl}\}~.
\end{equation}
\end{subequations}
Similarly, all other brackets can be computed.

\printbibliography

@book{stephani_exact_sol,
    place={Cambridge},
    edition={2},
    series={Cambridge Monographs on Mathematical Physics},
    title={Exact Solutions of Einstein's Field Equations},
    DOI={10.1017/CBO9780511535185},
    publisher={Cambridge University Press},
    author={Stephani, H. and Kramer, D. and MacCallum, M. and Hoenselaers, C. and Herlt, E.},
    year={2003},
    collection={Cambridge Monographs on Mathematical Physics}
}

@book{chandrasekharbook,
  title={The Mathematical Theory of Black Holes},
  author={Chandrasekhar, S.},
  isbn={9780198503705},
  lccn={93181092},
  series={International series of monographs on physics},
  url={https://books.google.com/books?id=LBOVcrzFfhsC},
  year={1998},
  publisher={Clarendon Press}
}

@article{LocalToroidalXanthapolous,
 ISSN = {00804630},
 URL = {http://www.jstor.org/stable/2397383},
 author = {B. C. Xanthopoulos},
 journal = {Proceedings of the Royal Society of London. Series A, Mathematical and Physical Sciences},
 number = {1794},
 pages = {117--131},
 publisher = {The Royal Society},
 title = {Local Toroidal Black Holes That are Static and Axisymmetric},
 urldate = {2023-09-06},
 volume = {388},
 year = {1983}
}

@article{GerochHartle,
    author = {Geroch, R. and Hartle, J. B.},
    title = "{Distorted black holes}",
    journal = {Journal of Mathematical Physics},
    volume = {23},
    number = {4},
    pages = {680-692},
    year = {1982},
    month = {04},
    issn = {0022-2488},
    doi = {10.1063/1.525384},
    url = {https://doi.org/10.1063/1.525384},
    eprint = {https://pubs.aip.org/aip/jmp/article-pdf/23/4/680/8152148/680\_1\_online.pdf},
}

@article{ashtekar2014,
      title={Geometry and Physics of Null Infinity}, 
      author={Ashtekar, A.},
      year={2015},
      eprint={1409.1800},
      archivePrefix={arXiv},
      primaryClass={gr-qc},
      url={https://arxiv.org/abs/1409.1800}, 
}

@article{bondi1962,
 author = {Bondi, H. and Burg, M. G. J. van der and Metzner, A. W. K.},
 doi = {10.1098/rspa.1962.0161},
 eprint = { },
 institution = { },
 issn = {0080-4630},
 journal = {Proceedings of the Royal Society of London. Series A. Mathematical and Physical Sciences},
 number = {1336},
 pages = {21--52},
 pmid = { },
 publisher = { },
 title = {{Gravitational waves in general relativity, VII. Waves from axi-symmetric isolated system}},
 url = { },
 volume = {269},
 year = {1962}
}

@Inbook{Geroch1977,
author="Geroch, R.",
editor="Esposito, F. Paul
and Witten, Louis",
title="Asymptotic Structure of Space-Time",
bookTitle="Asymptotic Structure of Space-Time",
year="1977",
publisher="Springer US",
address="Boston, MA",
pages="1--105",
isbn="978-1-4684-2343-3",
doi="10.1007/978-1-4684-2343-3_1",
url="https://doi.org/10.1007/978-1-4684-2343-3_1"
}

@article{mccarthy1972_structure,
 author = {McCarthy, P. J.},
 doi = {10.1063/1.1665917},
 eprint = { },
 institution = { },
 issn = {0022-2488},
 journal = {Journal of Mathematical Physics},
 number = {11},
 pages = {1837--1842},
 pmid = { },
 publisher = { },
 title = {Structure of the Bondi‐Metzner‐Sachs Group},
 url = { },
 volume = {13},
 year = {1972}
}

@article{sachs1962_waves,
 author = {Sachs, R. K.},
 doi = {10.1098/rspa.1962.0206},
 eprint = { },
 institution = { },
 issn = {0080-4630},
 journal = {Proceedings of the Royal Society of London. Series A. Mathematical and Physical Sciences},
 number = {1340},
 pages = {103--126},
 pmid = { },
 publisher = { },
 title = {{Gravitational waves in general relativity VIII. Waves in asymptotically flat space-time}},
 url = { },
 volume = {270},
 year = {1962}
}

@article{sachs1962_asym,
 author = {Sachs, R. K.},
 doi = {10.1103/physrev.128.2851},
 eprint = { },
 institution = { },
 issn = {0031-899X},
 journal = {Physical Review},
 number = {6},
 pages = {2851--2864},
 pmid = { },
 publisher = { },
 title = {{Asymptotic Symmetries in Gravitational Theory}},
 url = { },
 volume = {128},
 year = {1962}
}

@article{Anabalon2021,
  title={Closer look at black hole pair creation},
  author={Anabal{\'o}n, A. and Brenner, S. and Giribet, G. and Montecchio, L.},
  journal={Physical Review D},
  volume={104},
  number={2},
  pages={024044},
  year={2021},
  publisher={APS}
}

@article{Giribet2023,
    author = "Giribet, G. and Laurnagaray, J. and Schmied, P.",
    title = "{Probing the near-horizon geometry of black rings}",
    eprint = "2304.14461",
    archivePrefix = "arXiv",
    primaryClass = "hep-th",
    doi = "10.1103/PhysRevD.108.024061",
    journal = "Phys. Rev. D",
    volume = "108",
    number = "2",
    pages = "024061",
    year = "2023"
}

@article{Barnich-Troessaert,
    author = "Barnich, G. and Troessaert, C.",
    title = "{Aspects of the BMS/CFT correspondence}",
    eprint = "1001.1541",
    archivePrefix = "arXiv",
    primaryClass = "hep-th",
    reportNumber = "ULB-TH-09-28",
    doi = "10.1007/JHEP05(2010)062",
    journal = "JHEP",
    volume = "05",
    pages = "062",
    year = "2010"
}

@article{Donnay:2016,
    author = "Donnay, L. and Giribet, G. and Gonz\'alez, H. A. and Pino, M.",
    title = "{Extended Symmetries at the Black Hole Horizon}",
    eprint = "1607.05703",
    archivePrefix = "arXiv",
    primaryClass = "hep-th",
    doi = "10.1007/JHEP09(2016)100",
    journal = "JHEP",
    volume = "09",
    pages = "100",
    year = "2016"
}

@article{Donnay:2015,
    author = "Donnay, L. and Giribet, G. and Gonzalez, H. A. and Pino, M.",
    title = "{Supertranslations and Superrotations at the Black Hole Horizon}",
    eprint = "1511.08687",
    archivePrefix = "arXiv",
    primaryClass = "hep-th",
    doi = "10.1103/PhysRevLett.116.091101",
    journal = "Phys. Rev. Lett.",
    volume = "116",
    number = "9",
    pages = "091101",
    year = "2016"
}

@article{Donnay:2019,
    author = "Donnay, L. and Giribet, G.",
    title = "{Cosmological horizons, Noether charges and entropy}",
    eprint = "1903.09271",
    archivePrefix = "arXiv",
    primaryClass = "hep-th",
    doi = "10.1088/1361-6382/ab2e42",
    journal = "Class. Quant. Grav.",
    volume = "36",
    number = "16",
    pages = "165005",
    year = "2019"
}

@article{Hawking:2016,
    author = "Hawking, S. W. and Perry, M. J. and Strominger, A.",
    title = "{Superrotation Charge and Supertranslation Hair on Black Holes}",
    eprint = "1611.09175",
    archivePrefix = "arXiv",
    primaryClass = "hep-th",
    doi = "10.1007/JHEP05(2017)161",
    journal = "JHEP",
    volume = "05",
    pages = "161",
    year = "2017"
}

@misc{Hawking:2015,
      title={The Information Paradox for Black Holes}, 
      author={S. W. Hawking},
      year={2015},
      eprint="1509.01147",
      archivePrefix="arXiv",
      primaryClass={hep-th},
      url={https://arxiv.org/abs/1509.01147},
}

@article{PriceThorne,
  title = {Membrane viewpoint on black holes: Properties and evolution of the stretched horizon},
  author = {Price, R. H. and Thorne, K. S.},
  journal = {Phys. Rev. D},
  volume = {33},
  issue = {4},
  pages = {915--941},
  numpages = {0},
  year = {1986},
  month = {2},
  publisher = {American Physical Society},
  doi = {10.1103/PhysRevD.33.915},
  url = {https://link.aps.org/doi/10.1103/PhysRevD.33.915}
}

@article{Grumiller-2020,
  title = {Spacetime Structure near Generic Horizons and Soft Hair},
  author = {Grumiller, D. and P\'erez, A. and Sheikh-Jabbari, M. M. and Troncoso, R. and Zwikel, C.},
  journal = {Phys. Rev. Lett.},
  volume = {124},
  issue = {4},
  pages = {041601},
  numpages = {7},
  year = {2020},
  month = {01},
  publisher = {American Physical Society},
  doi = {10.1103/PhysRevLett.124.041601},
  url = {https://link.aps.org/doi/10.1103/PhysRevLett.124.041601}
}

@book{Grumiller-Jabbari-Book,
  title={Black Hole Physics: From Collapse to Evaporation},
  author={Grumiller, D. and Sheikh-Jabbari, M.M.},
  isbn={9783031103438},
  series={Graduate Texts in Physics},
  url={https://link.springer.com/book/10.1007/978-3-031-10343-8},
  doi={10.1007/978-3-031-10343-8},
  year={2022},
  publisher={Springer International Publishing}
}

@article{Wald-Zoupas,
  title = {General definition of ``conserved quantities'' in general relativity and other theories of gravity},
  author = {Wald, R. M. and Zoupas, A.},
  journal = {Phys. Rev. D},
  volume = {61},
  issue = {8},
  pages = {084027},
  numpages = {16},
  year = {2000},
  month = {03},
  publisher = {American Physical Society},
  doi = {10.1103/PhysRevD.61.084027},
  url = {https://link.aps.org/doi/10.1103/PhysRevD.61.084027}
}

@article{Barnich_Brandt_2002, 
title={Covariant theory of asymptotic symmetries, conservation laws and central charges},
volume={633}, 
ISSN={0550-3213}, 
DOI={https://doi.org/10.1016/S0550-3213(02)00251-1}, 
abstractNote={Under suitable assumptions on the boundary conditions, it is shown that there is a bijective correspondence between equivalence classes of asymptotic reducibility parameters and asymptotically conserved (n−2)-forms in the context of Lagrangian gauge theories. The asymptotic reducibility parameters can be interpreted as asymptotic Killing vector fields of the background, with asymptotic behaviour determined by a new dynamical condition. A universal formula for asymptotically conserved (n−2)-forms in terms of the reducibility parameters is derived. Sufficient conditions for finiteness of the charges built out of the asymptotically conserved (n−2)-forms and for the existence of a Lie algebra g among equivalence classes of asymptotic reducibility parameters are given. The representation of g in terms of the charges may be centrally extended. An explicit and covariant formula for the central charges is constructed. They are shown to be 2-cocycles on the Lie algebra g. The general considerations and formulas are applied to electrodynamics, Yang–Mills theory and Einstein gravity.}, 
number={1}, 
journal={Nuclear Physics B}, 
author={Barnich, G. and Brandt, F.}, 
year={2002}, 
pages={3–82} 
}

@article{Regge_Teitelboim_1974, title={Role of surface integrals in the Hamiltonian formulation of general relativity}, volume={88}, ISSN={0003-4916}, DOI={https://doi.org/10.1016/0003-4916(74)90404-7}, abstractNote={It is shown that if the phase space of general relativity is defined so as to contain the trajectories representing solutions of the equations of motion then, for asymptotically flat spaces, the Hamiltonian does not vanish but its value is given rather by a nonzero surface integral. If the deformations of the surface on which the state is defined are restricted so that the surface moves asymptotically parallel to itself in the time direction, then the surface integral gives directly the energy of the system, prior to fixing the coordinates or solving the constraints. Under more general conditions (when asymptotic Poincaré transformations are allowed) the surface integrals giving the total momentum and angular momentum also contribute to the Hamiltonian. These quantities are also identified without reference to a particular fixation of the coordinates. When coordinate conditions are imposed the associated reduced Hamiltonian is unambiguously obtained by introducing the solutions of the constraints into the surface integral giving the numerical value of the unreduced Hamiltonian. In the present treatment there are therefore no divergences that cease to be divergences after coordinate conditions are imposed. The procedure of reduction of the Hamiltonian is explicity carried out for two cases: (a) Maximal slicing, (b) ADM coordinate conditions. A Hamiltonian formalism which is manifestly covariant under Poincaré transformations at infinity is presented. In such a formalism the ten independent variables describing the asymptotic location of the surface are introduced, together with corresponding conjugate momenta, as new canonical variables in the same footing with the gij, πij. In this context one may fix the coordinates in the “interior” but still leave open the possibility of making asymptotic Poincaré transformations. In that case all ten generators of the Poincaré group are obtained by inserting the solution of the constraints into corresponding surface integrals.}, number={1}, journal={Annals of Physics}, author={Regge, T. and Teitelboim, C.}, year={1974}, pages={286–318} 
}

@article{Barnich:2006,
    author = "Barnich, G. and Compere, G.",
    title = "{Classical central extension for asymptotic symmetries at null infinity in three spacetime dimensions}",
    eprint = "gr-qc/0610130",
    archivePrefix = "arXiv",
    reportNumber = "ULB-TH-06-08",
    doi = "10.1088/0264-9381/24/5/F01",
    journal = "Class. Quant. Grav.",
    volume = "24",
    pages = "F15--F23",
    year = "2007"
}

@article{Penrose:1965,
    author = "Penrose, R.",
    title = "{Zero rest mass fields including gravitation: Asymptotic behavior}",
    doi = "10.1098/rspa.1965.0058",
    journal = "Proc. Roy. Soc. Lond. A",
    volume = "284",
    pages = "159",
    year = "1965"
}

@article{Newman-Penrose-1968,
    author = "Newman, E. T. and Penrose, R.",
    title = "{New conservation laws for zero rest-mass fields in asymptotically flat space-time}",
    doi = "10.1098/rspa.1968.0112",
    journal = "Proc. Roy. Soc. Lond. A",
    volume = "305",
    pages = "175--204",
    year = "1968"
}

@article{Barnich_2011,
   title={BMS charge algebra},
   volume={2011},
   ISSN={1029-8479},
   url={http://dx.doi.org/10.1007/JHEP12(2011)105},
   DOI={10.1007/jhep12(2011)105},
   number={12},
   journal={Journal of High Energy Physics},
   publisher={Springer Science and Business Media LLC},
   author={Barnich, G. and Troessaert, C.},
   year={2011},
   month=dec 
}

@article{Penrose:1962,
    author = "Penrose, R.",
    title = "{Asymptotic properties of fields and space-times}",
    doi = "10.1103/PhysRevLett.10.66",
    journal = "Phys. Rev. Lett.",
    volume = "10",
    pages = "66--68",
    year = "1963"
}

@article{McCarthy-74,
  title = {Physical Significance of the Topology of the Bondi-Metzner-Sachs Group},
  author = {Crampin, M. and McCarthy, P. J.},
  journal = {Phys. Rev. Lett.},
  volume = {33},
  issue = {9},
  pages = {547--550},
  numpages = {0},
  year = {1974},
  month = {08},
  publisher = {American Physical Society},
  doi = {10.1103/PhysRevLett.33.547},
  url = {https://link.aps.org/doi/10.1103/PhysRevLett.33.547}
}

@article{Newman1965,
  title={A Possible Connexion between the Gravitational Field and Elementary Particle Physics},
  author={E. T. Newman},
  journal={Nature},
  year={1965},
  volume={206},
  pages={811-812},
  url={https://api.semanticscholar.org/CorpusID:4220116}
}

@article{Strominger-Zhiboedov:2014,
    author = "Strominger, A. and Zhiboedov, A.",
    title = "{Gravitational Memory, BMS Supertranslations and Soft Theorems}",
    eprint = "1411.5745",
    archivePrefix = "arXiv",
    primaryClass = "hep-th",
    doi = "10.1007/JHEP01(2016)086",
    journal = "JHEP",
    volume = "01",
    pages = "086",
    year = "2016"
}

@article{Adami_2020a,
   title={T-Witts from the horizon},
   volume={2020},
   ISSN={1029-8479},
   url={http://dx.doi.org/10.1007/JHEP04(2020)128},
   DOI={10.1007/jhep04(2020)128},
   number={4},
   journal={Journal of High Energy Physics},
   publisher={Springer Science and Business Media LLC},
   author={Adami, H. and Grumiller, D. and Sadeghian, S. and Sheikh-Jabbari, M.M. and Zwikel, C.},
   year={2020},
   month=apr 
}

@article{Adami_2020b,
   title={Symmetries at null boundaries: two and three dimensional gravity cases},
   volume={2020},
   ISSN={1029-8479},
   url={http://dx.doi.org/10.1007/JHEP10(2020)107},
   DOI={10.1007/jhep10(2020)107},
   number={10},
   journal={Journal of High Energy Physics},
   publisher={Springer Science and Business Media LLC},
   author={Adami, H. and Sheikh-Jabbari, M.M. and Taghiloo, V. and Yavartanoo, H. and Zwikel, C.},
   year={2020},
   month=oct 
}

@article{Adami_2021b,
   title={Null boundary phase space: slicings, news and memory},
   volume={2021},
   ISSN={1029-8479},
   url={http://dx.doi.org/10.1007/JHEP11(2021)155},
   DOI={10.1007/jhep11(2021)155},
   number={11},
   journal={Journal of High Energy Physics},
   publisher={Springer Science and Business Media LLC},
   author={Adami, H. and Grumiller, D. and Sheikh-Jabbari, M. M. and Taghiloo, V. and Yavartanoo, H. and Zwikel, C.},
   year={2021},
   month=nov 
}

@article{Adami_2022,
   title={Null surface thermodynamics},
   volume={105},
   ISSN={2470-0029},
   url={http://dx.doi.org/10.1103/PhysRevD.105.066004},
   DOI={10.1103/physrevd.105.066004},
   number={6},
   journal={Physical Review D},
   publisher={American Physical Society (APS)},
   author={Adami, H. and Sheikh-Jabbari, M. M. and Taghiloo, V. and Yavartanoo, H.},
   year={2022},
   month=mar 
}

@article{Brill_1997,
   title={Thermodynamics of (3+1)-dimensional black holes with toroidal or higher genus horizons},
   volume={56},
   ISSN={1089-4918},
   url={http://dx.doi.org/10.1103/PhysRevD.56.3600},
   DOI={10.1103/physrevd.56.3600},
   number={6},
   journal={Physical Review D},
   publisher={American Physical Society (APS)},
   author={Brill, Dieter R. and Louko, Jorma and Peldán, Peter},
   year={1997},
   month=sep, 
   pages={3600–3610} 
}

@ARTICLE{Doroshkevich-Zeldovich-65,
       author = {{Doroshkevich}, A.~G. and {Zel'dovich}, Ya. B. and {Novikov}, I.~D.},
        title = "{Gravitational Collapse of Non-Symmetric and Rotating Bodies}",
      journal = {Zhurnal Eksperimentalnoi i Teoreticheskoi Fiziki},
         year = 1965,
        month = dec,
       volume = {49},
        pages = {170},
       adsurl = {https://ui.adsabs.harvard.edu/abs/1965ZhETF...49.170D}
}

@article{Safari-Jabbari:2019,
   title={BMS4 algebra, its stability and deformations},
   volume={2019},
   ISSN={1029-8479},
   url={http://dx.doi.org/10.1007/JHEP04(2019)068},
   DOI={10.1007/jhep04(2019)068},
   number={4},
   journal={Journal of High Energy Physics},
   publisher={Springer Science and Business Media LLC},
   author={Safari, H. R. and Sheikh-Jabbari, M. M.},
   year={2019},
   month=apr 
}

@article{Hawking:2016a,
  title = {Soft Hair on Black Holes},
  author = {Hawking, S. W. and Perry, M. J. and Strominger, A.},
  journal = {Phys. Rev. Lett.},
  volume = {116},
  issue = {23},
  pages = {231301},
  numpages = {9},
  year = {2016},
  month = jun,
  publisher = {American Physical Society},
  doi = {10.1103/PhysRevLett.116.231301},
  url = {https://link.aps.org/doi/10.1103/PhysRevLett.116.231301}
}

@article{Afshar:2017,
   title={Soft hairy horizons in three spacetime dimensions},
   volume={95},
   ISSN={2470-0029},
   url={http://dx.doi.org/10.1103/PhysRevD.95.106005},
   DOI={10.1103/physrevd.95.106005},
   number={10},
   journal={Physical Review D},
   publisher={American Physical Society (APS)},
   author={Afshar, H. and Grumiller, D. and Merbis, W. and Perez, A. and Tempo, D. and Troncoso, R.},
   year={2017},
   month=may 
}

@article{Akbar-Gibbons-2003,
   title={Ricci-flat metrics with U(1) action and the Dirichlet boundary-value problem in Riemannian quantum gravity and isoperimetric inequalities},
   volume={20},
   ISSN={1361-6382},
   url={http://dx.doi.org/10.1088/0264-9381/20/9/314},
   DOI={10.1088/0264-9381/20/9/314},
   number={9},
   journal={Classical and Quantum Gravity},
   publisher={IOP Publishing},
   author={Akbar, M. M. and Gibbons, G. W.},
   year={2003},
   month=apr, 
   pages={1787–1822} 
}

@article{York-Brown:1992,
    author = "Brown, J. D. and York, Jr., J. W.",
    title = "{Quasilocal energy and conserved charges derived from the gravitational action}",
    eprint = "gr-qc/9209012",
    archivePrefix = "arXiv",
    reportNumber = "IFP-423-UNC, TAR-009-UNC",
    doi = "10.1103/PhysRevD.47.1407",
    journal = "Phys. Rev. D",
    volume = "47",
    pages = "1407--1419",
    year = "1993"
}

@article{Akbar:2010,
  title = {Schwarzschild--anti-de Sitter black holes within isothermal cavity: Thermodynamics, phase transitions, and the Dirichlet problem},
  author = {Akbar, M. M.},
  journal = {Phys. Rev. D},
  volume = {82},
  issue = {6},
  pages = {064001},
  numpages = {14},
  year = {2010},
  month = sep,
  publisher = {American Physical Society},
  doi = {10.1103/PhysRevD.82.064001},
  url = {https://link.aps.org/doi/10.1103/PhysRevD.82.064001}
}

@book{Gibbons-Hawking-Book,
author = {Gibbons, G. W. and Hawking, S. W.},
title = {Euclidean Quantum Gravity},
publisher = {WORLD SCIENTIFIC},
year = {1993},
doi = {10.1142/1301},
address = {},
edition   = {},
URL = {https://www.worldscientific.com/doi/abs/10.1142/1301},
eprint = {https://www.worldscientific.com/doi/pdf/10.1142/1301}
}

@article{Carlip:2002,
    author = "Carlip, S.",
    title = "{Near horizon conformal symmetry and black hole entropy}",
    eprint = "gr-qc/0203001",
    archivePrefix = "arXiv",
    reportNumber = "UCD-02-02",
    doi = "10.1103/PhysRevLett.88.241301",
    journal = "Phys. Rev. Lett.",
    volume = "88",
    pages = "241301",
    year = "2002"
}

@article{Hotta_2001,
   title={Diffeomorphism on the horizon as an asymptotic isometry of the Schwarzschild black hole},
   volume={18},
   ISSN={1361-6382},
   url={http://dx.doi.org/10.1088/0264-9381/18/10/301},
   DOI={10.1088/0264-9381/18/10/301},
   number={10},
   journal={Classical and Quantum Gravity},
   publisher={IOP Publishing},
   author={Hotta, M. and Sasaki, K. and Sasaki, T.},
   year={2001},
   month=apr, 
   pages={1823–1834} 
}

@article{Carlip_1995,
   title={Statistical mechanics of the (2+1)-dimensional black hole},
   volume={51},
   ISSN={0556-2821},
   url={http://dx.doi.org/10.1103/PhysRevD.51.632},
   DOI={10.1103/physrevd.51.632},
   number={2},
   journal={Physical Review D},
   publisher={American Physical Society (APS)},
   author={Carlip, S.},
   year={1995},
   month=jan, 
   pages={632–637} 
}

@article{Carlip_1999,
   title={Entropy from conformal field theory at Killing horizons},
   volume={16},
   ISSN={1361-6382},
   url={http://dx.doi.org/10.1088/0264-9381/16/10/322},
   DOI={10.1088/0264-9381/16/10/322},
   number={10},
   journal={Classical and Quantum Gravity},
   publisher={IOP Publishing},
   author={Carlip, S.},
   year={1999},
   month=sep, 
   pages={3327–3348} 
}
\end{document}